\documentclass[
aps,prl,
reprint,
groupedaddress]{revtex4-1}

\usepackage[utf8x]{inputenc}
\usepackage{graphicx}

\begin{document}


\title{Endotaxial Si nanolines in Si(001):H}



\author{F. Bianco}
\affiliation{Department of Condensed Matter Physics, NCCR MaNEP,
University of Geneva, 24 Quai Ernest-Ansermet, 1211 Geneva 4,
Switzerland}

\author{J.H.G. Owen}
\affiliation{Department of Condensed Matter Physics, NCCR MaNEP,
University of Geneva, 24 Quai Ernest-Ansermet, 1211 Geneva 4,
Switzerland}

\author{S. A. Köster}
\affiliation{Department of Condensed Matter Physics, NCCR MaNEP,
University of Geneva, 24 Quai Ernest-Ansermet, 1211 Geneva 4,
Switzerland}

\author{D. Mazur}
\affiliation{Department of Condensed Matter Physics, NCCR MaNEP,
University of Geneva, 24 Quai Ernest-Ansermet, 1211 Geneva 4,
Switzerland}

\author{D. R. Bowler}
\affiliation{Department of Physics \& Astronomy, University College
London, Gower St, London WC1E~6BT, UK} \affiliation{London Centre
for Nanotechnology, 17--19 Gordon St, London WC1H~0AH, UK}

\author{Ch. Renner}
\email[]{christoph.renner@unige.ch}
\affiliation{Department of Condensed Matter Physics, NCCR MaNEP,
University of Geneva, 24 Quai Ernest-Ansermet, 1211 Geneva 4,
Switzerland}


\date{\today}

\begin{abstract}
We present a detailed study of the structural and electronic properties of a
self-assembled silicon nanoline embedded in the H-terminated silicon (001)
surface, known as the Haiku stripe. The nanoline is a perfectly straight and
defect free endotaxial structure of huge aspect ratio; it can grow micrometre
long at a constant width of exactly four Si dimers (1.54nm). Another remarkable
property is its capacity to be exposed to air without suffering any degradation.
The nanoline grows independently of any step edges at tunable densities, from
isolated nanolines to a dense array of nanolines. In addition to these unique
structural characteristics, scanning tunnelling microscopy and density
functional theory reveal a one-dimensional state confined along the Haiku core.
This nanoline is a promising candidate for the long sought after electronic
solid-state one-dimensional model system to explore the fascinating quantum
properties emerging in such reduced dimensionality.
\end{abstract}


\maketitle

\section{Introduction}
There is tremendous scientific and technological interest in physical systems of
reduced dimensionality.~\cite{Bowler04} Nanowires down to single atom chains are
of particular interest. The Fermi-liquid theory describing conventional bulk
conductors breaks down in such one-dimensional (1D) systems, which are predicted
to behave remarkably differently. The concept of a single particle becomes
obsolete and one has to switch to a description of collective spin and charge
excitation.~\cite{Giamarchi2004} However, the unusual properties of the
non-Fermi liquid counterpart, the Tomanaga-Luttinger liquid,~\cite{Luttinger65}
remain largely untested by experiment due to the lack of suitable practical
models for electronic systems.

The experimental quest for exotic quantum phenomena in reduced dimensions such
as quantized conductance,~\cite{Agrait03} charge density waves~\cite{Gruener94}
and Wigner crystals,~\cite{Schulz93} Peierls distortions~\cite{Peierls01} and
spin-charge separation~\cite{Zacher98} has triggered numerous developments in
the fabrication of nanostructures over the last decade.~\cite{Nitzan03,Barth05}
Optical and electron-beam lithography as well as nanoimprint techniques all
suffer from physical limitations and low fidelity on nanometer length scales.
Scanning probes enable the fabrication of nanostructures down to single-atom
assemblies.~\cite{Lyding94,Foelsch04,Hirjibehedin2006} However, this is a
tedious and not scalable procedure limited to assemblies of few tens of atoms
only. Moreover, their properties correspond to a particle in a box and are not
suitable to probe, for example, transport in 1D.

A very attractive alternative capable of producing single-atom scale structures
in a very efficient and scalable manner is self-assembly. Semiconductor
surfaces, especially silicon, are among the favourite substrates to
self-assemble nanowires. This is motivated by their compatibility with the
current industrial information technology processing,~\cite{Teichert02} and with
state-of-the-art nanoscale processing techniques.~\cite{Ruess2008a} Another
reason is the energy gap at the Fermi level allowing to decouple the nanowire
and substrate electronic states, which is paramount to obtaining 1D isolated
nanowires on a bulk substrate. A range of metal atoms have been self-assembled
into atomic chains on vicinal Si(553) and Si(557)
surfaces,~\cite{Himpsel04,Oncel08} and on Si(111) and Ge(001)
terraces.~\cite{Schaefer08,Schaefer09} In the first instance, the nanolines
assemble along the step edges, hence their spacing (density) is set by the
vicinal angle. In the second instance, the nanolines form a dense array. The
metallic rare earth silicides comprise another interesting family of 1D systems
that has been a focus of attention for many years.~\cite{Owen06a} Their 1D
growth is due to anisotropic epitaxial strain induced by unidirectional lattice
mismatch.~\cite{Bowler04} YSi$_{2}$ nanowires have been found to exhibit 1D
states.~\cite{Zeng08b} However, the major shortcoming of these structures is
their non-constant width and their high reactivity with air.

Here we present a novel and very promising self-assembled silicon only nanoline
structure embedded in the flat terraces of Si(001). These endotaxial (i.e.
epitaxially grown \emph{into} the substrate)~\cite{George91} nanolines, obtained
through hydrogenation of precursor Bi nanolines,~\cite{Owen06a,Owen10} exhibit a
range of remarkable properties. Like other endotaxial structures in Si(001),
such as silicide nanowires,~\cite{Zhian04} they show large aspect ratios with
lengths that can exceed 1\,$\mu$m. Moreover, they offer striking advantages over
all other nanolines mentioned above: they are absolutely straight at a constant
width of exactly four Si dimers (1.54\,nm); they are atomically perfect and
nearly defect free; they are stable upon air exposure for at least 25\,minutes;
they grow on the flat terrace away from any step edges and their density can be
tuned from isolated and non interacting nanolines to dense arrays of possibly
interacting nanolines.

We shall first describe the fabrication of these endotaxial Si nanolines. Then,
we discuss their structural and electronic properties in detail based on
experimental data from scanning tunnelling microscopy (STM), X-ray photoelectron
spectroscopy (XPS), reflection high energy electron diffraction (RHEED) and
density functional theory (DFT) modelling.

\section{Methods}
Endotaxial Si nanolines on Si(001) were fabricated and studied in ultra high
vacuum (UHV – base pressure in the low $10^{-11}$\,mbar) using an Omicron low
temperature STM. The sample preparation chamber is equipped with hot and cold
stages to control the substrate temperature during nanoline growth, several
Knudsen cells and electron beam evaporation sources, a high efficiency hydrogen
cracker and an ion gun. The nanowire growth is monitored in real time using
RHEED, and we further utilise XPS for chemical characterization.

Boron-doped p-type Si(001) substrates with a resistivity of 0.1\,$\Omega$cm were
prepared using a standard \textit{ex-situ} and \textit{in-situ} cleaning
sequence. \textit{Ex-situ} cleaning starts with degreasing the substrates in
ultrasonic acetone followed by isopropanol baths for 5 minutes each. Each
substrate is then rinsed in purified water and cleaned in an oxidising mixture
of H$_2$O$_2$ and H$_2$SO$_4$ (1:1 by volume) for 30 seconds. This surface oxide
is subsequently removed by a 30 second HF (15 wt\%) dip. Finally, the above
oxidation step is repeated to regrow a thin protective oxide film and the
substrate is rinsed in purified water and dried with nitrogen gas.
\textit{In-situ} cleaning starts by degassing the Si(001) substrate for
typically 12 hours at 600$^{\circ}$C while keeping the pressure below
$1\times10^{-9}$\,mbar. To obtain a clean surface, it is then repeatedly
flash-annealed at 1200$^{\circ}$C while making sure the pressure is kept below
$2\times10^{-9}$\,mbar. This step is repeated until the substrate can be
flashed-annealed for 10–15\,s. The last flash-annealing is followed by a
controlled cool-down from 900$^{\circ}$C to the nanowire growth temperature of
570$^{\circ}$C in roughly 1 minute.

To obtain the precursor nanolines (Fig.\,\ref{fig:rheed}~(a)), bismuth is
evaporated
for 10 to 15 minutes onto the clean Si(001) surface kept at 570$^{\circ}$C from
a MBE Komponenten K-cell heated to 470$^{\circ}$C. After the deposition, the
sample is kept at 570$^{\circ}$C for 2 to 3 additional minutes, and then cooled
down to the hydrogenation temperature of 300$^{\circ}$C. The entire deposition
process is monitored in real time with RHEED. The characteristic arc in the
RHEED pattern (Fig.\,\ref{fig:rheed}~(b)) signalling the formation of Bi
nanolines
appears 6 to 8 minutes into the deposition process.

The Bi nanolines thus obtained are exposed to a hydrogen atomic beam from a MBE
Komponenten HABS high efficiency hydrogen cracker heated to
1600$^{\circ}$C.~\cite{Tschersich08} The sample temperature is set at
300$^{\circ}$C during the hydrogenation process. Hydrogen is fed into the UHV
chamber through the cracker itself and the flow rate is controlled by
maintaining a constant chamber pressure of $2.5\times10^{-7}$\,mbar. The Si(001)
surface and the Bi nanolines are exposed to the atomic hydrogen flux for
4–8\,minutes, corresponding to doses of 60--120\,L. We found that this amount of
hydrogen is sufficient to saturate all the Si dangling bonds to form a
monohydride surface.~\cite{Boland90} It also fully strips the bismuth off the
surface, saturating the exposed Si bonds with hydrogen but without affecting the
underlying one dimensional Haiku reconstruction.~\cite{Owen10} This results in a
1D endotaxial hydrogen saturated Si nanoline, called the Haiku
stripe,~\cite{Owen10} which is the subject of detailed investigations in the
following sections.

We performed DFT~\cite{Hohenberg64,Kohn65} calculations using a plane-wave
implementation within the generalised gradient approximation (GGA). All
calculations have been performed with the Vienna \emph{ab initio} simulation
package (VASP).~\cite{Kresse93,Kresse96} Ultrasoft
pseudopotentials~\cite{Vanderbilt90} are used for all the elements considered.
Convergence of forces and energy differences is achieved by selecting a
Monkhorst-pack~\cite{Monkhorst76} mesh of $4\times2\times1$ to sample the
Brillouin zone, and an energy cutoff of 200 eV.  We used a unit cell which was
one dimer row wide, ten dimers long and ten layers deep, with the bottom two
layers fixed and the bottom layer terminated with
hydrogen.~\cite{Rodriguez-Prieto09a,Rodriguez-Prieto09b}

\section{Results}

\subsection{Haiku stripes}

%
%
\begin{figure}
 \includegraphics[width=.9\linewidth]{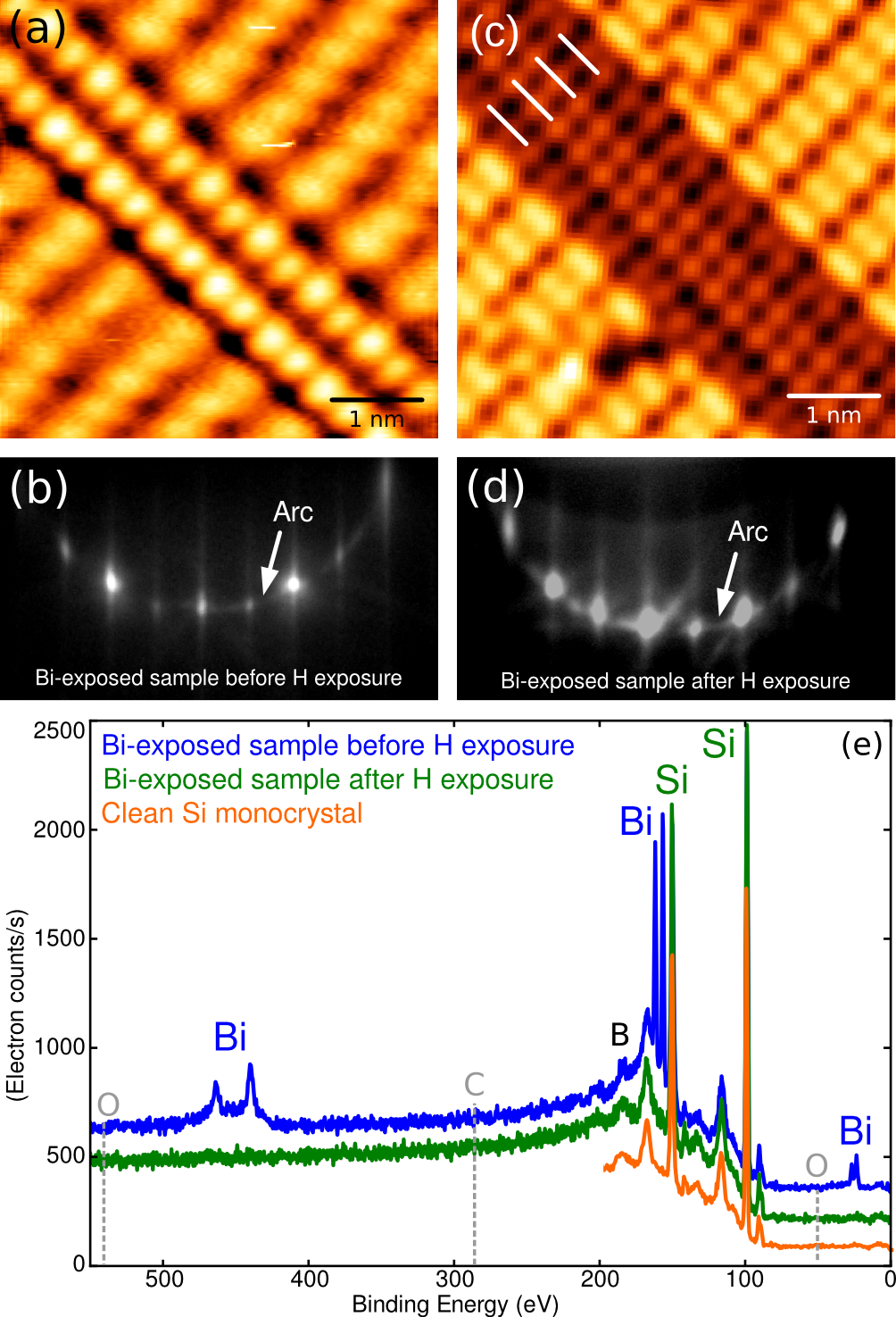}
 \caption{
(a) High resolution STM micrograph of a Bi nanoline on a clean Si(001)
background. (sample bias $V = -3.5$\,V, $I = 200$\,pA, $T = 77$\,K) (b) RHEED
pattern after Bi growth. The arc signifies the Bi nanoline and the bright dots
are due to the 2${\times}$1 reconstruction of the Si surface. (c) High
resolution STM micrograph of the Haiku stripe (4 Si dimers wide diagonal
structure) revealed after the hydrogen induced removal of the Bi atoms. ($V =
-2.8$\,V, $I = 200$\,pA, $T = 77$\,K) (d) RHEED pattern showing the
characteristic arc of a well-ordered 1D structure on the surface, the Haiku
stripe. (e) XPS spectra measured on the Si(001) surface at the different states
of sample preparation. The small peak at 188\, eV is due to B dopant atoms. No
peaks from contamination of C or O are resolved. The top curves are offset by
+100 and +200\,counts/s for clarity.}
 \label{fig:rheed}
\end{figure}

While the Bi nanolines appear to STM as a pair of Bi surface dimers, there is a
substantial reconstruction of the underlying Si lattice – a triangular core of
Si embedded in the top five layers of the Si substrate called the Haiku core.
The Bi nanolines, with the unique Haiku structure, have been found to be the
lowest energy reconstruction of Si(001):Bi.~\cite{Bowler02} In DFT calculations,
the Haiku structure is 0.36\,eV/Bi dimer more favourable than the next best
proposed structure,~\cite{Bowler02} and 0.25\,eV/Bi dimer more favourable than
one monolayer of Bi.~\cite{Owen02b} A remarkable result of our work is to find
that this structure remains stable in the hydrogenated Si(001) surface, despite
the removal of the Bi dimers. Both the Bi nanoline (Fig.\,\ref{fig:rheed}~(a,b))
and the Haiku stripe (Fig.\,\ref{fig:rheed}~(c,d)) manifest themselves as an arc
in the RHEED pattern measured along \textless110\textgreater{}, a direct
consequence of their 1D nature and direct evidence of the subsurface
reconstructed core. Annealing experiments on Haiku stripes showed that the arc
feature remains in RHEED up to a temperature of about 400$^{\circ}$C. This
contrasts with previous experiments and simulations on burial of the nanolines
by Si which found that the Haiku core structure spontaneously dissociated during
overgrowth by Si.~\cite{Sakata05} Fig.\,\ref{fig:rheed}~(e) shows XPS spectra of
the Bi-exposed surface before and after exposure to hydrogen. Characteristic Bi
peaks at 464, 440, 157, 159 and 25\,eV, which are clearly resolved after the Bi
nanoline growth, completely disappear upon hydrogenation of the Si(001) surface.
The XPS spectrum of the hydrogenated surface is indistinguishable from the
pristine Si(001) surface. No contaminant species is detected, except for a
minute B signal originating from the bulk dopant atoms. STM micrographs provide
further evidence that the Haiku stripe imaged in Fig.\,\ref{fig:rheed}~(c) is
not just a different appearance of the Bi nanoline due to particular imaging
conditions. The Haiku stripes appear dark independent of bias, which
unquestionably distinguishes them from the Bi nanolines whose STM image contrast
is always bright relative to a H-terminated surface.~\cite{Owen06a}

\begin{figure}
 \includegraphics[width=.9\linewidth]{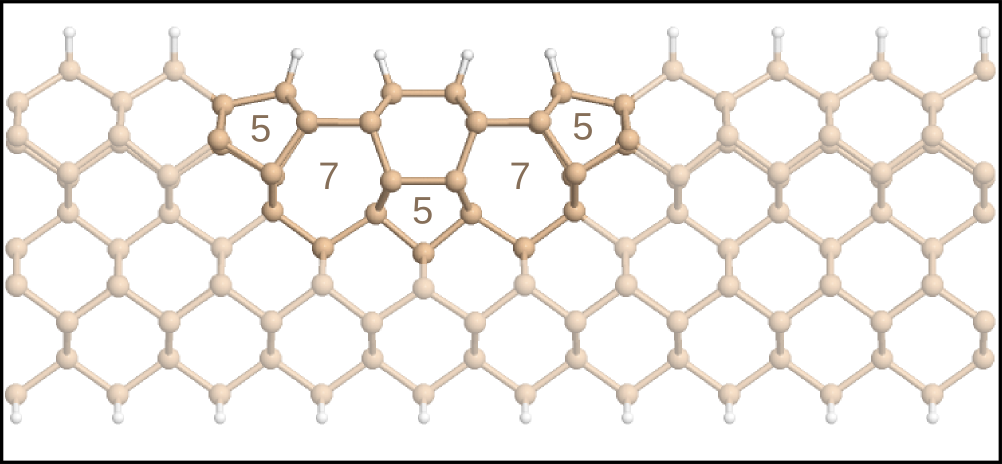}
 \caption{Ball-and-stick model of the DFT structure for the Haiku stripe showing
the 5- and 7- fold ring reconstruction of the Haiku core (cross section).
 }
 \label{fig:model}
\end{figure}

In Fig.\,\ref{fig:rheed}~(c) the STM micrograph of the hydrogen covered Haiku
core structure shows well resolved dark stripes. The dark stripes observed after
the Bi has been stripped off can be explained by a replacement of each Si-Bi
bond pair by two Si-H bonds.~\cite{Owen10} The proposed structure from DFT of
the resulting H saturated endotaxial Si nanoline is illustrated in the
ball-and-stick model of Fig.\,\ref{fig:model}. Its key feature is the Haiku core
– a major reconstruction of the silicon extending 5 layers below the surface and
composed of 5- and 7-fold Si rings. The four rows of dots indicated by short
solid white lines in the STM micrograph of the Haiku stripe
(Fig.\,\ref{fig:rheed}~(c)) correspond to the four H atoms spanning the Haiku
stripe over exactly 4 Si dimers (1.54\,nm). These rows appear 60 to 70\,pm
deeper than the surrounding Si(001):H surface, in quantitative agreement with
the value expected from the relaxed structure in the DFT simulation
(Fig.\,\ref{fig:model}), and in good agreement with the depth of similar
P-terminated 5-7-5 structures.\cite{McMahon2006a}

The above result suggests that Bi can be fully stripped off from the Bi
nanolines. This is opposite to previous experiments on hydrogenation of Bi
nanolines which have reported that the nanolines were inert,~\cite{Owen06a}
except for a recent indication of Bi nanoline damage with a very large dose
(1000\,L H$_{2}$) by Wang et al.~\cite{Wang08} We explain this controversy on
the basis of the different hydrogenation setups used. All the previous
experiments were done using a hot filament, whereas we are using a high
efficiency hydrogen cracker. The geometry of the H-cracker exposes the sample to
a much greater density of hot hydrogen accelerated towards the sample. The gas
flux impacts the sample at an angle of 25$^{\circ}$. By contrast, a hot filament
sends the gas atoms and molecules in all directions, thus reducing the flux
density. The exact mechanism by which H replaces the Bi atoms is not understood
at present. One possibility is that hot uncracked H$_2$ molecules reach the
surface with sufficient kinetic energy to attack the Bi-Si bonds, thus exposing
Si dangling bonds which are quickly capped by H atoms. To test this idea, we
lowered the temperature of the H cracker, thus greatly reducing the flux of
atomic hydrogen. We observed no reduction of the Bi nanoline stripping rate,
suggesting that hot, high kinetic energy H$_2$ molecules hold the key to the
observed stripping effect.

%
%
Since the Si(001) surface is very reactive to water, we took special care to
reduce the exposure to impurity water. We carefully baked the gas line back to
the H-bottle, and further cryopumped the hydrogen gas used in the gas line. In
addition, we have used a dedicated cracker, degassed over many hours at
operating temperature (1600$^{\circ}$C). The chamber pressure with the cracker
at 1600$^{\circ}$C is $8\times10^{-10}$\,mbar. The cracker has a much
higher efficiency than a hot filament.~\cite{Tschersich08} This allows us to
achieve a saturation H coverage with a very small amount of H$_{2}$ gas
(60-120L), thereby greatly reducing the exposure to any impurities. To achieve
saturation of the surface with a hot filament typically H$_{2}$ doses of
500-1000\,L are required~\cite{Hersam01} (15-20 minutes exposure). In our case
any possible carbon or oxygen on the surface after atomic H dosing was below the
XPS detection limit (Fig.\,\ref{fig:rheed}~(e)). Although we cannot measure the
water partial pressure directly, we believe that any impurity water has been
minimised, and that any effect on the nanoline will come from the exposure to
hydrogen.

%
%

%
%
\begin{figure}
 \includegraphics[width=.9\linewidth]{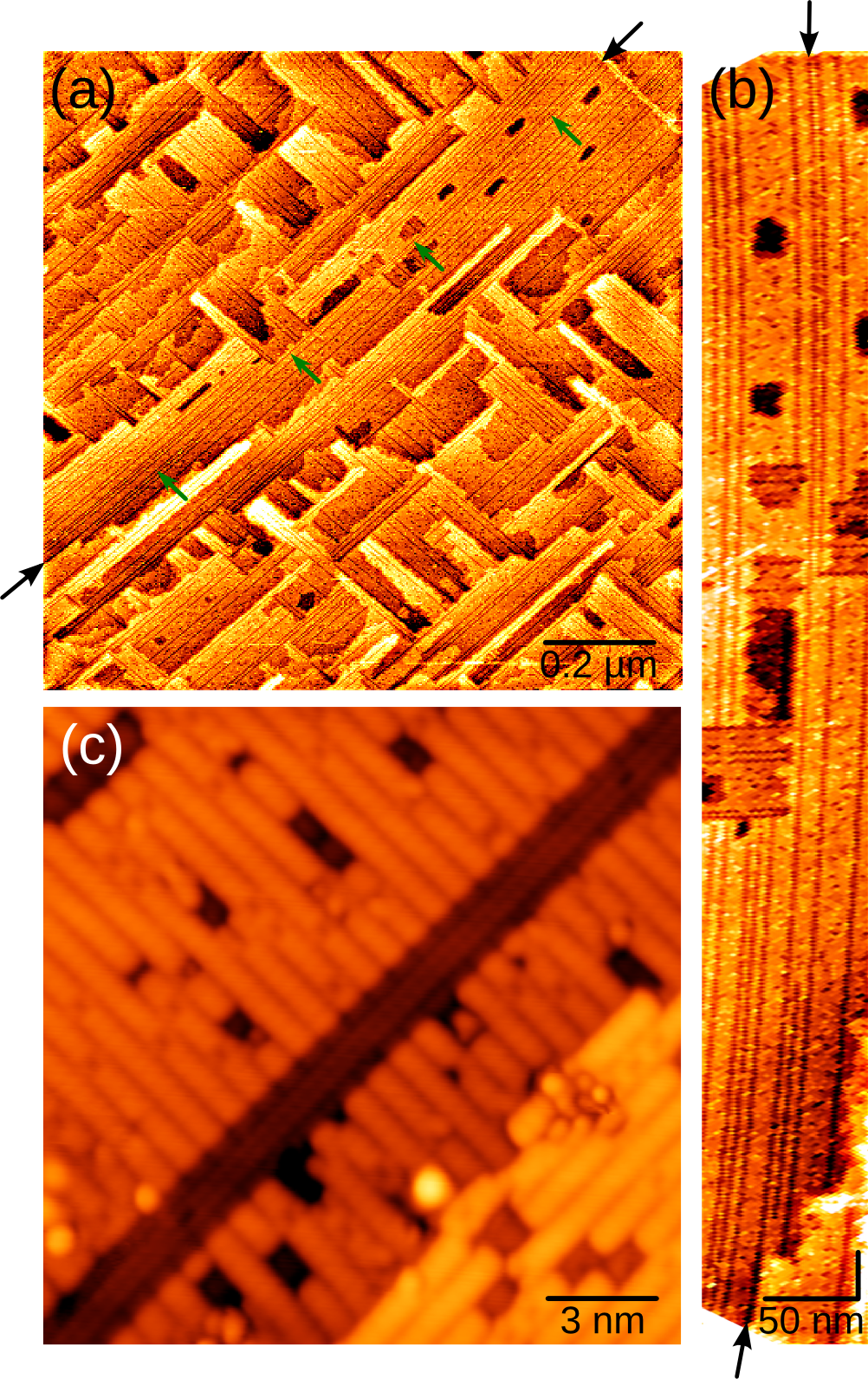}
 \caption{
(a) STM image of long Haiku stripes. The black arrows denote the start and
end points of a 1.3\,$\mu$m long stripe, while the green arrows are guides for
the eyes. ($V = -2.5$\,V, $I = 180$\,pA, $T = 77$\,K) (b) Magnification of the
1.3\,$\mu$m long stripe denoted by arrows in (a). For clarity the horizontal
axis has been stretched by a factor of two. (c) STM image taken after 25\,min
air exposure without any special treatment showing a Haiku stripe running
diagonally though the image. Neither the surface nor the Haiku stripe have been
attacked by reactants. The dark structures away from the stripe on the
substrate, also present on as grown surfaces, are either missing Si atoms or
places where background Bi has been extracted by H. ($V = -2.2$\,V, $I =
100$\,pA, $T = 77$\,K)
}
 \label{fig:air}
\end{figure}

Fig.\,\ref{fig:air}~(a) shows a large scale micrograph where many long stripes
(\textgreater{}0.2\,$\mu$m) can be seen. The longest stripe on this STM
micrograph, whose extremities are identified by black arrows in
Fig.\,\ref{fig:air}~(a,b), reaches 1.3\,$\mu$m. The terrace containing this
stripe is shown in more detail in Fig.\,\ref{fig:air}~(b).
Fig.\,\ref{fig:air}~(a) confirms that the Haiku stripes share a number of
features with the Bi nanolines: they are straight without any kinks and have a
fixed width, they are seen on terraces and terminate at defects and step edges.
These properties have been attributed to the Haiku core. Additionally, the
density of the Haiku stripes on the surface corresponds to the density of the
precursor Bi nanolines, and such stripes have never been seen after
hydrogenation of the clean Si(001) surface. These facts substantiate our
assumption that these lines result from hydrogenation of precursor Bi nanolines,
and the Haiku core is the correct model for this endotaxial Si-in-Si(001)
nanoline.

It has been known for many years that the Si(001) monohydride surface is stable
in air for up to 40\,hours.~\cite{Hersam01} Since the Si-H bond along the Haiku
stripes is of similar strength as the usual Si-H bond on the hydrogenated
Si(001) surface according to DFT, it is reasonable to expect air stability of
the stripes. We checked this hypothesis experimentally, and
Fig.\,\ref{fig:air}~(c) shows a Haiku stripe which is indeed still intact after
25\,min exposure to air. The sample was transferred to the load lock, which was
then slowly vented, while the sample was protected from high velocity molecules
by lying at an angle pointing away from the leak valve as recommended by
ref.~\onlinecite{Hersam01}. Then the sample was stored in air for 25\,min in a
simple plastic box. After the exposure to air, the sample was reinserted into
the load lock and transferred to the STM without any additional treatment.

The length of the Haiku stripes and their air stability offers a unique prospect
to contact them to electrodes enabling standard transport experiments to be
carried out. The air stability allows to consider common \emph{ex-situ}
semiconductor lithographic techniques unlike the precursor Bi nanolines, or
Rare-Earth silicide nanowires. In UHV, this can also be realised either by
depositing metal on top of the surface using shadow-mask technology, or by STM
lithography using the H layer as a mask. Recent developments in the latter
area\cite{Ruess04a} provide the ability to perform more complex contacting and
gating of the stripes on a few-nm length scale.

\subsection{Electronic structure}

STM is ideally suited to probe some of the quantum phenomena emerging in 1D
systems, in particular charge density waves, local spin phases and van Hove
singularities. Bias dependent STM imaging is a very effective way to distinguish
between purely electronic contrast due to, for example, localised states, and
structural contrast reflecting atomic positions. In the following, we compare
topographic images measured at different sample bias voltages and tunnelling
current with DFT simulations. The experiment and model are in excellent
agreement, and reveal the existence of an extended 1D state at the surface along
the Haiku core.

%
%
\begin{figure}
 \includegraphics[width=.9\linewidth]{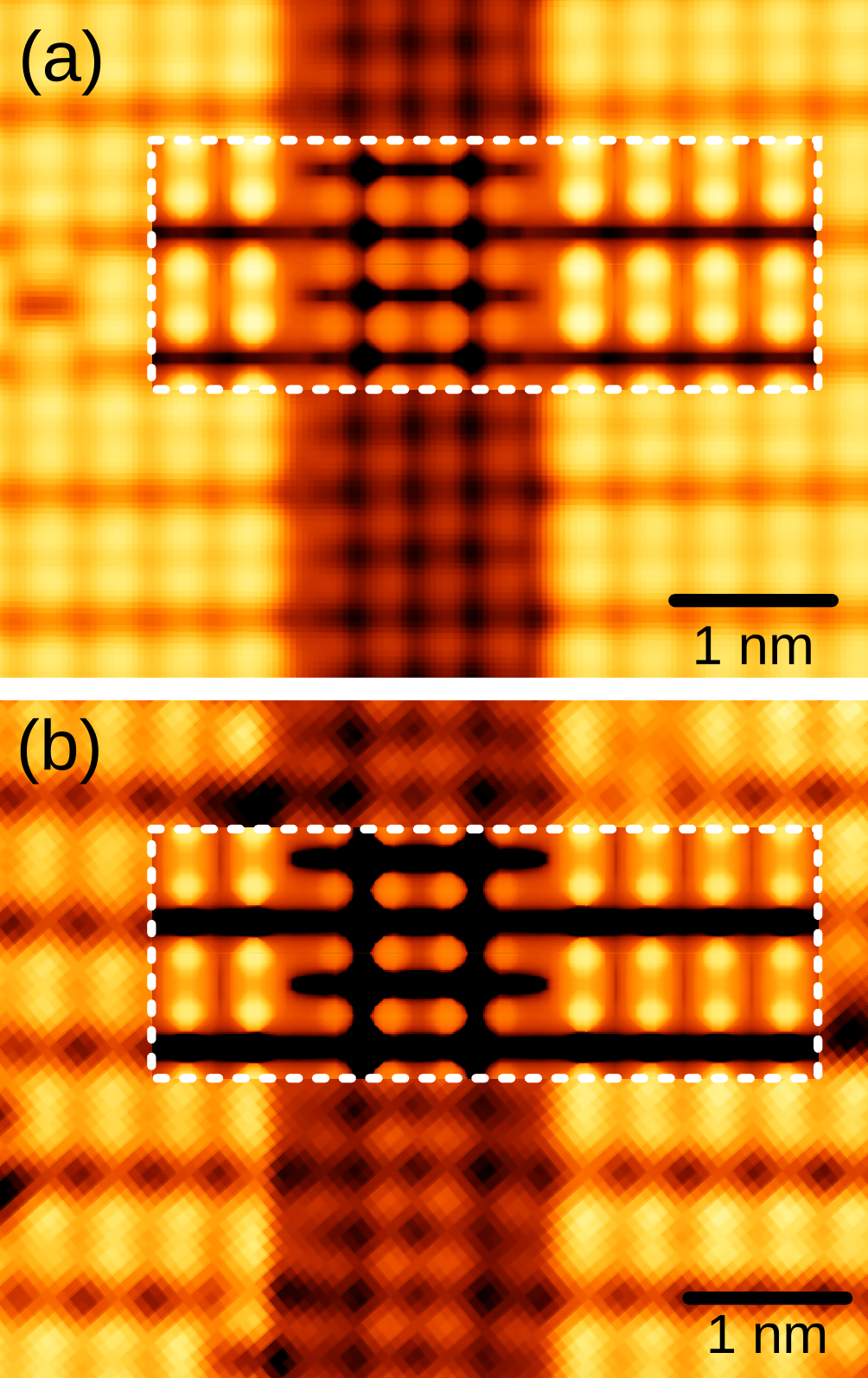}
 \caption{
Filled state high resolution STM micrographs of the Haiku stripe with the
corresponding DFT simulations shown within the dashed rectangles. (a) Sample
bias $V = -2.5$\,V, $I = 80$\,pA, $T = 77$\,K, DFT at -1.5\,eV for low current.
(b) $V = -2.8$\,V, $I = 200$\,pA, $T = 77$\,K, DFT at -1.5\,eV for high current.
}
 \label{fig:fstate}
\end{figure}

High resolution STM micrographs (positive sample bias) reveal the Haiku stripe
as a row of four dark atoms. In Fig.\,\ref{fig:fstate}~(a) recorded at a low
current, they appear regularly spaced and all at the same elevation about 70\,pm
lower (darker color) than the surrounding Si(001) surface. At a higher
tunnelling current (Fig.\,\ref{fig:fstate}~(b)), the middle two atomic sites
appear raised by 5–10\,pm compared to the two outer atomic sites of the Haiku
stripe. STM simulations (inset outlined by dashed rectangles in
Fig.\,\ref{fig:fstate}~(a,b)) based on the DFT relaxed structure of the Haiku
stripe (Fig.\,\ref{fig:model}) perfectly reproduce all of the above
characteristics. They were calculated for the same tunnelling bias, but
different electron density isosurfaces to simulate different tunnelling
currents. Fig.\,\ref{fig:model} shows that the four atomic sites spanning the
Haiku stripe are physically at the same elevation, suggesting that the height
contrast within the four atoms wide stripes is of purely electronic origin. The
DFT simulation further reproduces very accurately the lateral position of the
atomic sites, with a clear shift of the two outer sites towards the edges of the
Haiku stripe when observed at the higher tunnelling current
(Fig.\,\ref{fig:fstate}~(b)).

%
%
\begin{figure*}
 \includegraphics[width=.9\linewidth]{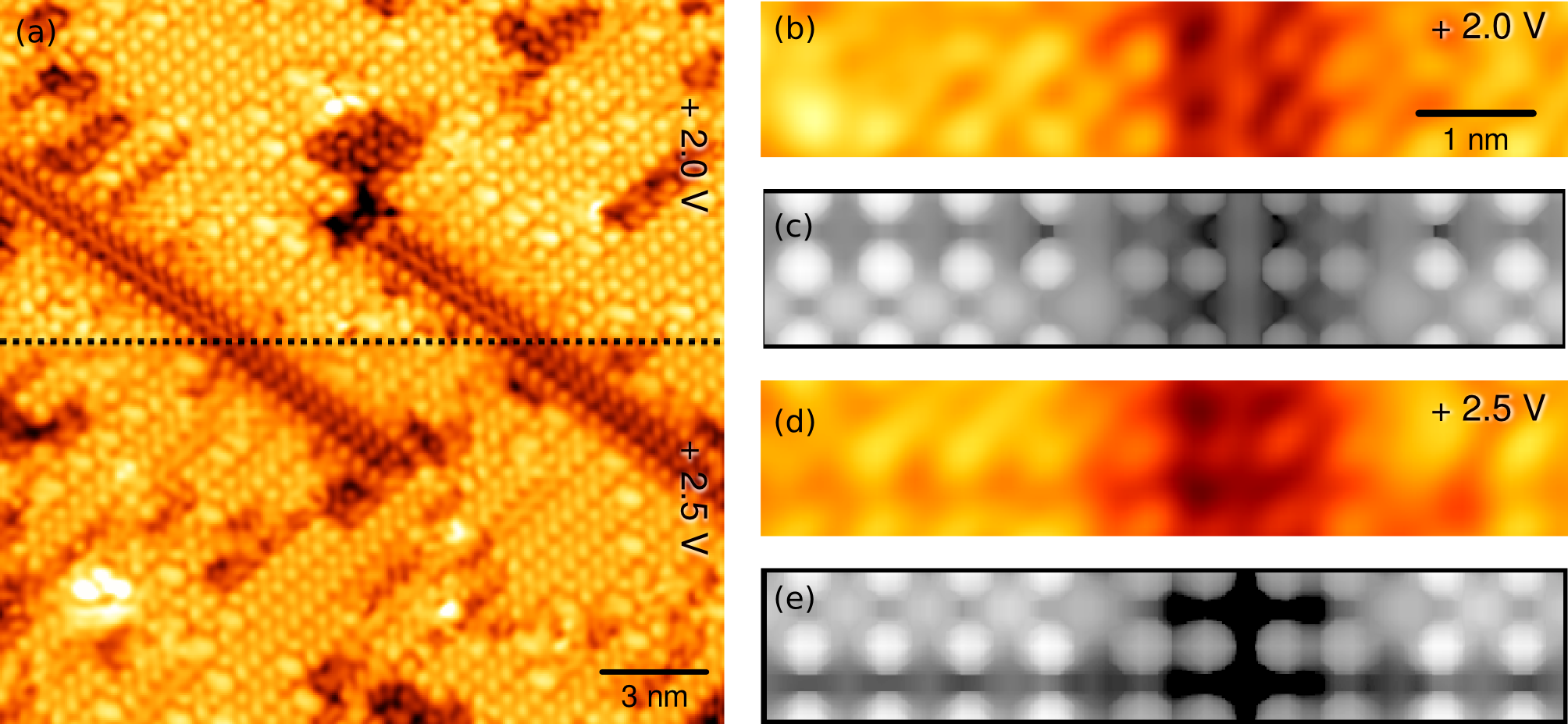}
 \caption{
(a) Positive sample bias (empty states) STM micrographs of two Haiku stripes.
Note the 1D state observed in the upper half of the image recorded at
$V=+2.0$\,V, which disappears when increasing the bias to $+2.5$\,V in the lower
half of the micrograph ($ I = 150$\,pA, $T = 77$\,K). (b) A magnification of the
central state of the Haiku stripe seen at $V = +2.0$\,V and (c) corresponding
DFT model calculated for an energy of +1.0\,eV. (d) A magnification of the Haiku
stripe at $V = +2.5$\,V and (e) the corresponding DFT model calculated for an
energy of +2.5\,eV.
}
 \label{fig:estate}
\end{figure*}

Imaging the Si(001) at a positive tunnelling bias, which probes the empty states
of the sample also reveals the four atomic sites across the Haiku stripe as
shown in Fig.\,\ref{fig:estate}. The top half and the bottom half in panel (a)
were taken one after the other with the same tip while only changing the bias
magnitude. In contrast to the negative bias images (Fig.\,\ref{fig:fstate}), the
two central atoms along the Haiku stripe appear shifted away from each other,
while the outer two are shifted towards the centre of the structure
(Fig.\,\ref{fig:estate}). The central atoms now appear lower than the outer ones
— the apparent height difference is reversed compared to the filled state
images. All these features illustrated in the magnified STM image sections
(Fig.\,\ref{fig:estate}~(b,d)) are perfectly reproduced in the corresponding DFT
simulations (Fig.\,\ref{fig:estate}~(c,e)).

The most remarkable feature in the empty state STM images is a bright linear
feature extending along the centre of the Haiku stripe
(Fig.\,\ref{fig:estate}~(a) top). It is only seen in a narrow energy window
around +2.0\,V (Fig.\,\ref{fig:estate}~(a)), and does not match any atomic
position in the Haiku model (Fig.\,\ref{fig:model}). Note that the background
Si(001) is the same in the top and bottom sections of the micrograph shown in
Fig.\,\ref{fig:estate}~(a). Again, all these characteristics are perfectly
reproduced by DFT; the simulation at +1.0\,V (Fig.\,\ref{fig:estate}~(c)) shows
a linear electronic feature in the middle of the structure, whereas it is absent
from the simulation at higher energy (Fig.\,\ref{fig:estate}~(e)).

%
%
\begin{figure}
 \includegraphics[width=.95\linewidth]{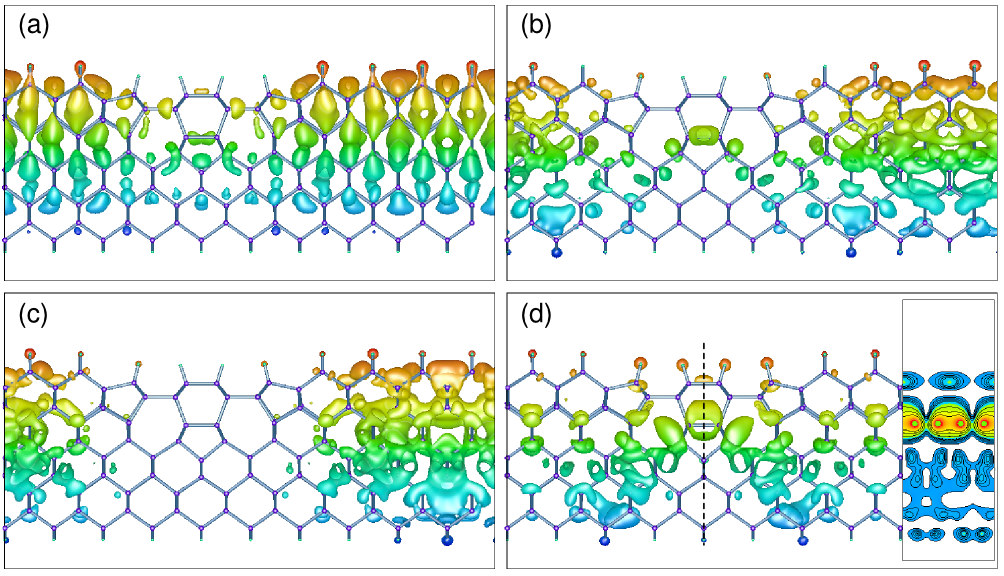}
 \caption{(a)–(d) Contour plots of constant charge densities on the Haiku stripe
cross section for the four bands above the Conduction Band Minimum (CBM). The
isosurface colour represents the height within the structure. (a) The CBM (b)
+0.02\,eV (c) +0.04\,eV (d) +0.10\,eV above the CBM.  In (d), the charge density
is located in the middle of the Haiku stripe. A cut through the charge density
taken at the position of the dashed line reveals a delocalised state along the
Haiku stripe. On the cut, the colour corresponds to the charge densities: the
blue is a low density and red is a high charge density.}
 \label{fig:densities}
\end{figure}

In order to identify the origin of this 1D state, we computed the charge
densities in the vicinity of a Haiku stripe. Fig.\,\ref{fig:densities} shows DFT
isosurfaces of constant charge density for four bands starting at the conduction
band minimum (CBM) towards higher energy states corresponding to the states
probed primarily by STM at positive sample bias. The isosurfaces demonstrate the
substantially different electronic structure below the surface of the Haiku
core, compared to pure Si. The first three bands above the lowest unoccupied
molecular orbital (LUMO) show little to no charge density within the Haiku core
and none of them are continuous along the stripes. On the other hand, the fourth
band in Fig.\,\ref{fig:densities}~(d) has its maximal charge density in the
middle of the Si reconstruction. A cross-section through the charge density
along the centre of the stripe (Fig.\,\ref{fig:densities}~(d), inset) shows that
this fourth band is delocalised along the endotaxial nanoline, in excellent
agreement with the STM observation (Fig.\,\ref{fig:estate}~(a), top).

\subsection{Discussion}

%
%
The Haiku stripes are extremely high aspect ratio Si-only 1D structures of
unprecedented perfection. They offer a range of very attractive features towards
the experimental exploration of 1D quantum physics. There is the experimental
opportunity to attach contacts, which is paramount to study transport in a truly
isolated 1D model system. The prime avenue will be the 1D state confined to the
Haiku core, but delocalized along it, which we have identified in STM and
confirmed by DFT. Because it is located just above the minimum of the conduction
band, one challenge will be to dope it to a suitable level. Local removal of the
H atoms on the top of the stripe, followed by adsorption of PH$_{3}$
(ref.~\onlinecite{Ruess04a}) could lead to local doping of this state, for
example. The physical and electronic structure of the stripes is well
understood, so that experiments can enjoy strong theoretical support, while the
micrometre length gives access to the infinite length regime addressed by
theory. Moreover, the degree of interaction between neighbouring nanowires is
tunable. Adjustable nanoline density is achieved by means of the growth
temperature and Bi flux during the self-assembly of the Bi nanoline precursor.
This is distinctly different from other self-assembled nano-scale wires, whose
density can either not be adjusted\cite{Wang2004a} or is set by the step density
of a vicinal surface.~\cite{Himpsel04}

As well as the intrinsic properties of the Haiku stripes, they can also be used
as a template for self assembling 1D chains of other atomic species. Previous
experiments have shown templating properties for the Bi nanolines.\cite{Owen06b}
According to modelling, the stripes host favourable adsorption sites for a range
of metal atoms. For example, Cu may diffuse to subsurface sites, forming
encapsulated single-atom chains.~\cite{Rodriguez-Prieto09b} We are investigating
the possibility that Mn atoms may also move into subsurface sites, producing
atomic chains with strong magnetic properties. The H-termination of the Haiku
stripes offers the exciting possibility that such subsurface metal atomic chains
would be stable in air.

Looking more broadly in the direction of future device fabrication, the H
termination of these nanolines and consequent air stability, their micrometre
length and their stability to an elevated temperature make these nanolines
compatible with a wide variety of semiconductor processing technologies.

\section{Conclusions}
We have described the properties of a novel endotaxial Si nanoline in Si(001)
identified as the Haiku stripe. Its core holds a delocalised state just above
the conduction band minimum, which combined with the length and the perfection
of the Haiku stripe makes it a very promising and appealing structure to explore
quantum properties emerging in 1D. Moreover, the stability to air and to higher
temperature, which we have demonstrated here, offer exciting experimental
opportunities. The detailed matching of DFT simulations and STM micrographs
confirms that the structure of the stripes is very well understood and settles
the Haiku core model on firm ground. Thus the discovery of this unique Si-in-Si
nanoline opens possibilities for exploring 1D physics and has a great potential
for the fabrication of nanoscale electronic devices on a technologically
relevant substrate.

\begin{acknowledgments}
The authors are very grateful to G. Manfrini for his expert technical support,
and to A. Rodriguez-Prieto for scientific discussions. We thank the Swiss
National Science Foundation for financial support through the National Center of
Competence in Research MaNEP (Materials with Novel Electronic Properties) and
division II. D.R. Bowler was supported by the Royal Society, and acknowledges
computational facilities in the London Centre for Nanotechnology.
\end{acknowledgments}

%

\end{document}